# Microwave coherent spectroscopy of ultracold thulium atoms

D.A. Pershin [1,2], V.V. Yaroshenko[3], V.V. Tsyganok [1,4,5], V.A. Khlebnikov [1], E.T. Davletov[1,4], D.V. Shaykin[1,4], E.R. Gadylshin[1,4], I.S. Cojocaru[1,2,6], E.L. Svechnikov[3], P.V. Kapitanova[3], A.V. Akimov[1,2,6]

[1]*Russian Quantum Center, Business Center "Ural", 100A Novaya Street Skolkovo, Moscow, 143025, Russia*

[2]*PN Lebedev Institute RAS, Leninsky Prospekt 53, Moscow, 119991, Russia*

[3]*Faculty of Physics and Engineering, ITMO University, 197101 Saint Petersburg, Russia*

[4]*Moscow Institute of Physics and Technology, Institutskii pereulok 9, Dolgoprudny, Moscow Region 141701, Russia*

[5] *National University of Science and Technology MISIS, Leninsky Prospekt 4, Moscow, 119049, Russia,*

[6]*Texas A&M University, TAMU 4242, College Station, TX 77843, USA*

email: akimov@physics.tamu.edu

## I. ABSTRACT

Recently, the thulium atom was cooled down to the Bose-Einstein condensation temperature, thus opening a pathway to quantum simulation with this atom. However, successful simulations require instruments to control and readout states of the atom as well as the ability to control the interaction between either different species or different states of the same type of species. In this paper, we provide an experimental demonstration of high-fidelity (over 93%) manipulation of the ground state magnetic sublevels of thulium, which utilizes a simple and efficient design of a microwave (MW) antenna. The coherence time and dephasing rate of the energetically highest hyperfine level of the ground state were also examined.

## II. INTRODUCTION

Ultracold atoms have become a solid platform for quantum simulations [1–3]. Manipulation of the interatomic interactions in such a simulator can be routinely achieved using so-called Fano-Feshbach resonances [4]. In particular, thulium, having the single bosonic isotope $^{169}$Tm, was recently cooled down to the Bose-Einstein condensation (BEC) temperature [5]. It has relatively large orbital angular momentum $L=3$ and magnetic dipole moment $\mu = 4\mu_B$ in the ground state along with a relatively simple level structure compared to other highly magnetic rare earth elements [6,7] and could thus be useful for quantum simulations. In the case of the thulium atom, similar to other rare earth elements [8–11], Fano-Feshbach resonances are accessible at low (Gauss level) magnetic fields [12].

For quantum simulations, the abilities to manipulate the ground state and to populate specific ground state components with high fidelity are of great importance. The atoms cooled down to nearly BEC temperature are usually polarized in the ground state (in the presence of a direct current (DC) external magnetic field); thus, there is a rather pure state with a single sublevel populated [13]. In the case of thulium, the ground state is $|F=4, m_F=-4\rangle$. A polarization purity in cold atomic gas of well over 95% is experimentally achievable [14]. In the presence of a small external magnetic field, population of other Zeeman sublevels could be achieved using direct radio frequency excitation of these transitions [15] via, for instance, resonant $\pi-$pulses [16]. This method is well established and has been developed into complicated composite pulses correcting for various errors in $\pi-$pulses [17–19] as well as the adiabatic rapid passage technique [20]. Nevertheless, the major difficulty with these methods is the necessity of a strong low-frequency field, which often needs to be delivered into a metallic vacuum volume. Additionally, this method is hard to implement in very low or zero magnetic fields. Thus, a Raman-type scheme is often more convenient for such experiments [21,22].

While a Raman-type population may be realized via optical levels, this approach may be quite demanding on the laser sources in the case of two levels that are very closely located in energy. Thus, utilization of the ground state hyperfine structure MW $F=4 \rightarrow F=3$ transition with frequency $\Delta_{hf}=1496.55$ MHz [23,24] seems to be a convenient approach (see Figure 1A). An additional advantage of this approach is the relatively long lifetime of the intermediate state, enabling cascade excitation instead of the commonly used stimulated adiabatic rapid passage. While this MW radiation still needs to be delivered into a metallic, in our case, vacuum volume, this task is solvable [25].

In this paper, we demonstrate manipulation of the ultracold thulium atom ground state using a Raman-like approach. The manipulation was performed using a specially developed MW antenna. The fidelity of transfer as well as the coherence properties of the intermediate hyperfine structure level was estimated.

## III. EXPERIMENTAL SETUP

Thulium atoms were cooled and trapped using a procedure described elsewhere [14,25,26]. The temperature of the cloud was set to $T \simeq 1.6 \pm 0.2 \ \mu K$, and the number of atoms in the crossed beam optical dipole trap (532 nm) [5] polarized in the $|F=4, m_F=-4\rangle$ substate was approximately $10^5$. Detection of the atomic cloud was performed via absorption imaging [27], described in detail in 0. The atom polarization was maintained with a vertical DC magnetic field $B = 4.09 \pm 0.04$ G (Figure 1C, storage).

In the presence of an external magnetic field, the ground state of $^{169}$Tm $4f^{13}(^2F^o)6s^2\ ^2F^o_{7/2}(I=1/2)$ ($I$ stands for nuclear spin) splits into Zeeman sublevels (Figure 1A,C), which are separated, if one neglects the quadratic Zeeman shift, by frequencies $g_F B \mu_B / h$, where $g_F$ stands for the Lande g-factor, $B$ is the magnetic field, $h$ is the Planck constant, and $\mu_B$ is the Bohr magneton. The g-factors of the hyperfine components of the ground state are $g_{F=4} = 0.999$ and $g_{F=3} = 1.284$ (see APPENDIX B). Since these g-factors are different, the magnetic dipole allowed transitions between hyperfine components of the ground state have different frequencies, thus allowing frequency selective addressing of specific transitions with an MW field.

The preparation procedures allow one to polarize atoms in the ground state $|F=4, m_F=-4\rangle$ along the external magnetic field [14]. Since the magnetic field creates quantization axes, magnetic dipole allowed transitions could be excited efficiently with circular or linear polarization of the MW field, thus populating any desired state. Experimentally, transitions with frequencies $f_{43}$ and $f_{33}$ were selected because of their reasonably large transition strengths (1/9 and 1/36, respectively, see Figure 1A). The transitions could thus be excited with circular and linear polarization, respectively. To address both these polarizations (and potentially a negative circular polarized transition as well), it was decided to use an MW source with a linear MW magnetic field component along the $x$ axis.

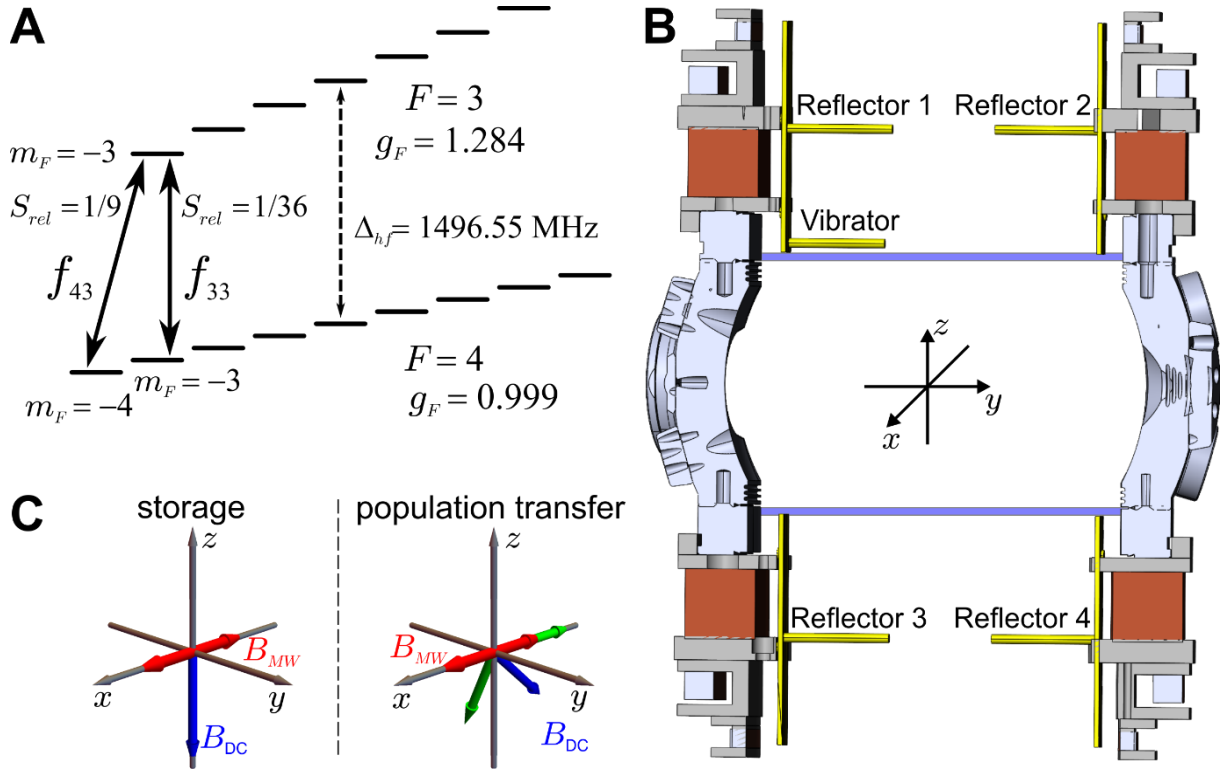

*Figure 1 A – level scheme of the thulium atom ground state. $F$ stands for the total atom momentum, $m_F$ for its projection onto the quantization axis (DC magnetic field), $S_{rel}$ for the relative strength of the transition, or square of the 3-j symbol of the transition, $f$ for the transition frequency, $g$ for the Lande g-factor, and $\Delta_{hf}$ for the hyperfine splitting frequency. B – vacuum chamber with MW antenna installed. C – orientations of the DC and MW magnetic fields used in the experiments. Green arrows indicate components of the DC magnetic field along the MW magnetic field direction and perpendicular to it.*

To address the transition between hyperfine levels of the ground state, an MW antenna, depicted in Figure 1B, was developed. In contrast to the previous version [25], this design has one active (vibrator) and 4 passive (reflectors) elements. The active element is a nonsymmetric vibrator that together with the reflector immediately below it can be considered as the simplest Yagi–Uda antenna [28]. The antenna elements are in the $yz$ plane, thus generating an $x$-polarized magnetic field. The combination of the bottom and top reflectors forms a cavity, which has a resonance frequency close to 1.5 GHz. The antenna has a bandwidth of 10 MHz for the $-10$ dBm level and a maximum magnetic field in the central region of the vacuum volume (see Figure 2). We note that in principle, one could try to use the vacuum chamber itself as a cavity, but unfortunately, its vacuum volume size does not allow excitation of the desired frequency.

Initial optimization of the $x$ component of the MW magnetic field was performed with CST MW Studio. The obtained $x$ component of the magnetic field distribution is presented in Figure 2. The antenna parameters optimized during the simulations are summarized in Table 1. One can see from Figure 2 that the magnetic field distribution is asymmetric in the $xy$ and $yz$ planes, which can be explained by the asymmetric field excitation. However, the maximum magnetic field is observed in the chamber center, reaching a value of 43 mG. One may conclude that the antenna efficiently utilizes the excitation power despite the presence of asymmetry. From the vector plot of the magnetic field distribution, one can see that the field has mainly a vertical component (x component). The gradient of the MW magnetic field in the center of the chamber does not exceed 1% per millimeter.

*Table 1 MW antenna parameters. The source is assumed to be 10 W at 1.5 GHz*

| Magnetic field intensity | 43 mG |
|---|---|
| Length of the active vibrator | 41.8 mm |
| Length of reflector 1 | 48.6 mm |
| Length of reflector 2 | 45.5 mm |
| Length of reflectors 3 and 4 | 45 mm |
| Distance between the vibrator and the 1$^{st}$ reflector | 60 mm |
| Diameter of vibrator/reflectors | 4 mm |
| Material | Brass |

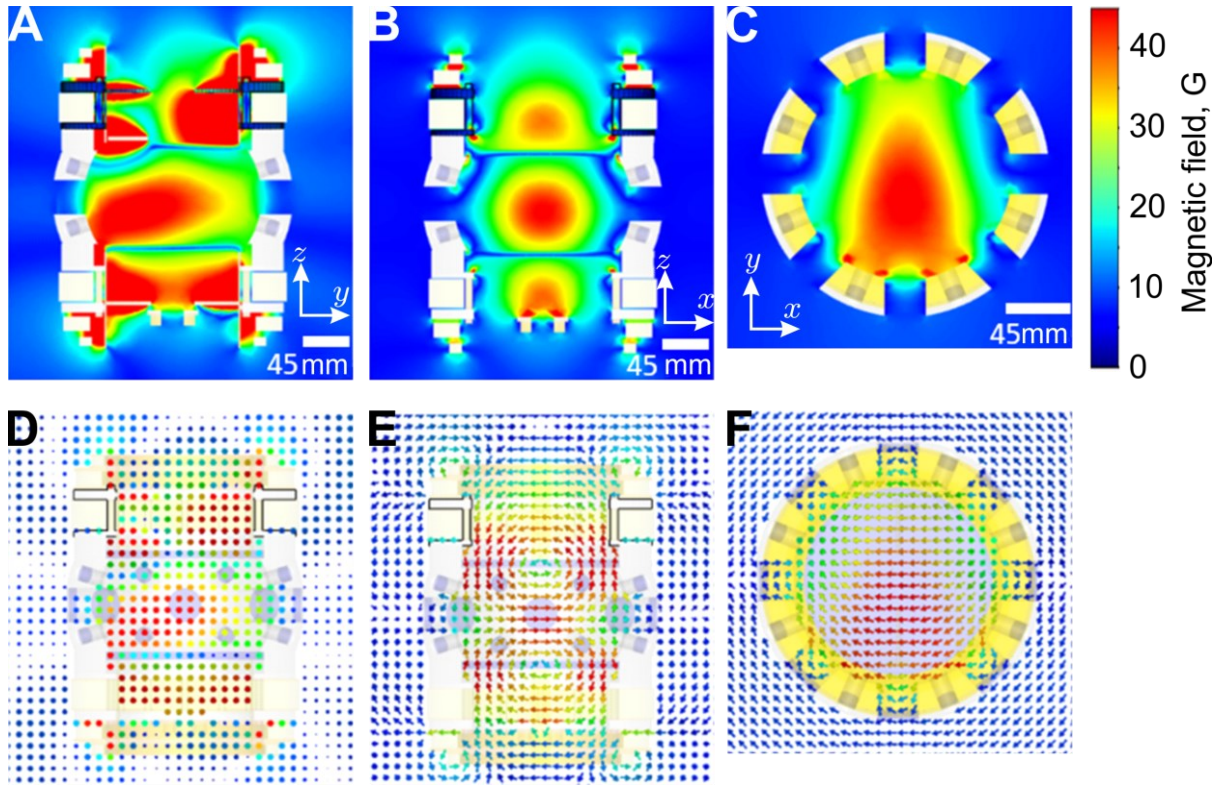

*Figure 2 Numerically calculated distribution of the $x$ component of the MW magnetic field in the vacuum chamber in the $A - yz$, $B - xz$, and $C - xy$ planes, and vector distribution of the MW magnetic field in the chamber in the $D - yz$, $E - xz$, and $F - xy$ planes.*

The feeding network of the MW antenna is illustrated in Figure 3A. To generate an MW signal, an SG 384 oscillator (Stanford Research Systems, SRS in the Figure) and the tracking generator of an HMS3010 spectrum analyzer (Rhode and Schwartz, R&S in the scheme) were used. The two generators were used to enable fast frequency switching, which was implemented using a ZASWA-2-50DR switch (Mini Circuits). The signal of the sources was modulated with another switch of the same type. The free output of the switch was loaded with 50 ohm. The modulated signal was then amplified with a ZHL-10W-2G+ amplifier (Mini Circuits, Amp in the figure) and fed to the antenna. The switches were controlled by a field-programmable gate array (National Instruments NI PCIe-6363, FPGA in Figure 3A) and a Stanford Research Systems DG 645 delay generator (Delay in the figure), thus allowing realization of the various pulse schemes demonstrated in Figure 3B.

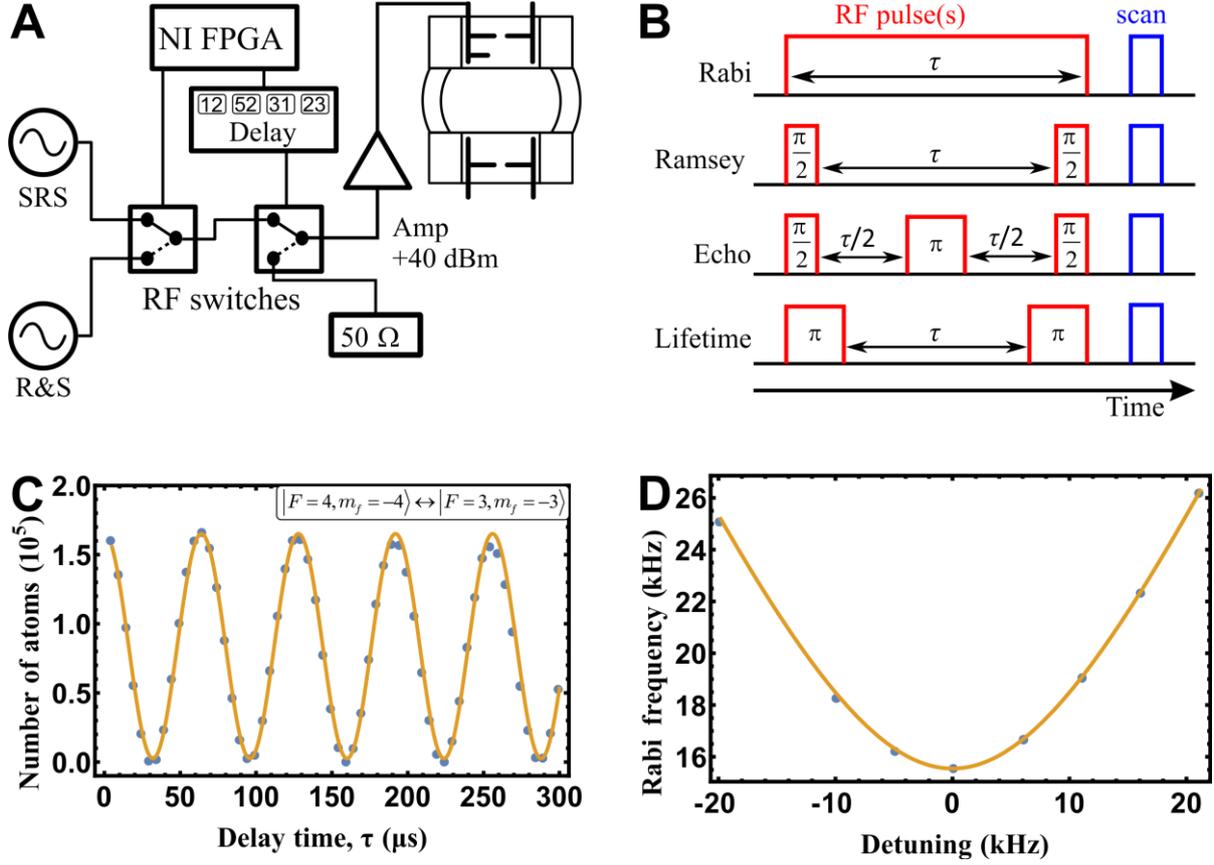

*Figure 3 A – diagram of the MW part of the experimental setup. SRS stands for Stanford Research Systems SG 384, R&S stands for Rhode and Schwartz HMS3010 (generator output), RF switches stands for Mini Circuits ZASWA-2-50DR, Delay stands for Stanford Research Systems DG645, Amp stands for Mini Circuits ZHL-10W-2G+ amplifier, and NI FPGA stands for National Instruments NI PCIe-6363. B – pulse sequences used for the experiment described in the paper. C – the data (dots) show the Rabi oscillations of the atomic ensemble for the transition with frequency $f_{43}$, and the solid line represents the fit function* (1) *with parameters* $\Omega = 2\pi \cdot 15630 \text{ Hz}$, $a = 0.81 \cdot 10^5$, *and* $b = 0.03 \cdot 10^5$. *D – the data (dots) show the generalized Rabi frequency versus detuning. The solid line represents the fit function* (2) *with parameters* $\Omega_{Rabi} = 2\pi \cdot 15550 \text{ Hz}$ *and* $f_{34} = 1497367190 \text{ Hz}$.

## IV.     RESULTS

To determine the magnitude of the experimentally achieved MW magnetic field, a Rabi oscillation sequence was used (Rabi in Figure 3B), thus allowing observation of the Rabi oscillation at the $f_{43}$ transition frequency (Figure 3C). The data were fitted with the following dependence of the $|F=4, m_F = -4\rangle$ level population $n_{44}$:

$$n_{44} = a\left(1 + \cos(\Omega\tau)\right) + b \tag{1}$$

where $a, b$ are fit parameters, $\tau$ is time and $\Omega$ is the Rabi frequency [29]. Since the Rabi frequency depends on detuning $\Delta$ as $\Omega = \sqrt{\Omega_{Rabi}^2 + \Delta^2}$, with $\Omega_{Rabi}$ being the resonance Rabi frequency, the detuning was chosen to be 0 by fitting the dependence of the Rabi frequency on the frequency $f$ applied to the transition (Figure 3D, see APPENDIX C) with the following function:

$$\Omega = \sqrt{\Omega_{Rabi}^2 + (f - f_{34})^2} \qquad (2)$$

For this experiment, the direction of the DC magnetic field was kept vertical, as it is set for atom loading and storage in the trap (Figure 1C). For such an orientation of the field, the atoms experience the maximum possible circular component of the MW field. The MW antenna was tuned within the experimental setup (see APPENDIX C) to reach the maximum allowed by the design Rabi frequency of $\Omega_{Rabi} = 2\pi \cdot 15.55$ kHz. The magnitude of the AC magnetic field was found to be $B = 21 \pm 1$ mG using the following equation:

$$\vec{B} \cdot \vec{\mu} = \hbar \Omega_{Rabi} \qquad (3)$$

where $\hbar$ is the reduced Planck constant and $\mu$ is the magnetic dipole moment of the transition (see APPENDIX C for details). The value measured this way $(51 \pm 3)\%$ is less than that obtained from the simulations, presented in Figure 2.

The fidelity of the population transfer from the $|F = 4, m_F = -4\rangle$ to $|F = 4, m_F = -3\rangle$ states strongly depends on the coherence properties of the atomic ensemble and achievable Rabi frequency. Thus, both the dephasing time $T_2^*$ and the coherence time $T_2$ for the transition with frequency $f_{43}$ were examined. This was done using well-established Ramsey [30] and Hahn echo [31] techniques. Figure 3B demonstrates the MW pulse sequences used for these experiments. Here, the temperature of the cloud was set to $T \simeq 3.6 \pm 0.2$ $\mu$K.

In the Ramsey experiment, atoms were exposed to two resonant MW pulses ("Ramsey" in Figure 3B). The first was a π/2-pulse to create a superposition of ground and excited states, and the second was a π/2 or 3π/2-pulse to transfer atoms to excited or ground states, respectively. The results of the two sequences were subtracted to measure the pure coherence-related decay signal. After the pulses, atoms were released, and their number was measured at 500 μs of expansion. The time between the two pulses $\tau$ was varied to observe dephasing of the atomic ensemble. The result is shown in Figure 4A; the difference is fitted perfectly with exponential decay $e^{-\tau/T_2^*}$ with dephasing time $T_2^* = 150 \pm 10$ μs.

The echo-type experiment has an additional π-pulse in the middle between the two pulses mentioned above ("Echo" in Figure 3B), thus allowing us to remove some dephasing of the atomic ensemble. The idea of such an experiment remains the same unless the roles of the second π/2 and 3π/2 pulses are switched. The result is shown in Figure 4B, and the difference is fitted with $e^{-(\tau/T_2)^\alpha}$ [32], providing an echo-related time of coherence of $T_2 = 500 \pm 20$ μs, with $\alpha = 1.9 \pm 0.2$.

The population decay from the $|F=3, m_F = -3\rangle$ state was measured at a magnetic field of 5.3 G, as indicated in Figure 4C. This measurement was done by applying a $\pi$-pulse for the transition with frequency $f_{43}$ followed by the same $\pi$-pulse after some delay time $\tau$, which was varied. After the second $\pi$-pulse, the trap was released. The measurement of the number of atoms was performed after 2 ms of free expansion of the atomic ensemble by absorption imaging. It is interesting that the population decay is not a single exponential but is rather well described by binary collision decay [33,34]. As a rule, the linear loss $\gamma$ of a dipole trap is ~1 s$^{-1}$, which allows us to fit the decay curve with only binary collisions:

$$N(t) = \frac{N_0}{1 + N_0 \beta^* t} \tag{4}$$

where $N_0$ is the initial number of atoms in the trap, and $\beta^* = \beta/V$, where $\beta$ is a binary collision coefficient and $V$ is the volume of the atomic cloud (see APPENDIX D). The fit parameter was found to be $\beta = 2.23^{+0.8}_{-0.5} \times 10^{-9}$ cm$^3$/s. The contribution of the binary collisions was faster than the linear decay rate, suggesting the presence of dipolar relaxation [35,36], a light-assisted process [33,37–39] or Feshbach resonance [4]. The population decay has a characteristic time of approximately 20 ms, which is much longer than the coherence time, indicating the presence of strong decoherence.

The relatively short coherence time of the hyperfine structure transitions limits how slow the MW pulses could be. Since the transition with frequency $f_{33}$ is much weaker than that with frequency $f_{43}$ (see Figure 1A), it is important to optimize the distribution of the MW field between the two transitions. Thus, to realize population transfer from the $|F=4, m_F = -4\rangle$ state to the $|F=4, m_F=-3\rangle$ state, the DC magnetic field (and, correspondingly, atomic polarization [14]) was tilted (Figure 1C population transfer, the magnetic field is directed along the vector $(-0.76(3), -0.28(1), -0.59(2))$ with a magnitude of $1.06\pm0.04\,\text{G}$) such that the Rabi frequencies for the weaker transition with the $f_{33}$ frequency (see Figure 1A) and stronger transition with the $f_{43}$ frequency were approximately the same: $2\pi \cdot (7120\pm20)\,\text{Hz}$ and $2\pi \cdot (9510\pm30)\,\text{Hz}$, respectively. The data were fitted with models:

$$\begin{aligned} n &= ae^{-\tau/T_1}\left(1\pm\cos(\Omega\tau)\right)+b \\ n_{44} &= ae^{-\tau/T_1}\left(1+\cos(\Omega\tau)\right)+b \end{aligned} \quad (5)$$

where + is selected for level $|F=4, m_F=-4\rangle$ and − for level $|F=4, m_F=-3\rangle$. Here, $n$ stands for the population of the corresponding level forming transitions with frequencies $f_{33}$ and $f_{43}$, $a$ and $\Omega$ represent the amplitude and frequency of the Rabi oscillations, parameter $T_1$ is the population decay time of the atoms in the trap, and $b$ represents the possible background. With these settings, Rabi oscillations for both transitions with $f_{33}$ and $f_{43}$ frequencies were observed by detecting the population of states with $F=4$ (see Figure 4D and 0). With these nearly equal Rabi frequencies, population transfer was realized with the sequence of two $\pi$ − pulses at $f_{43}$ and $f_{33}$ with durations of 67 μs and 55 μs, respectively, and a 73 μs delay in between. In addition, the trap was turned off during the second pulse as well as 70 μs before and 5 μs after it to reduce trap-related dephasing.

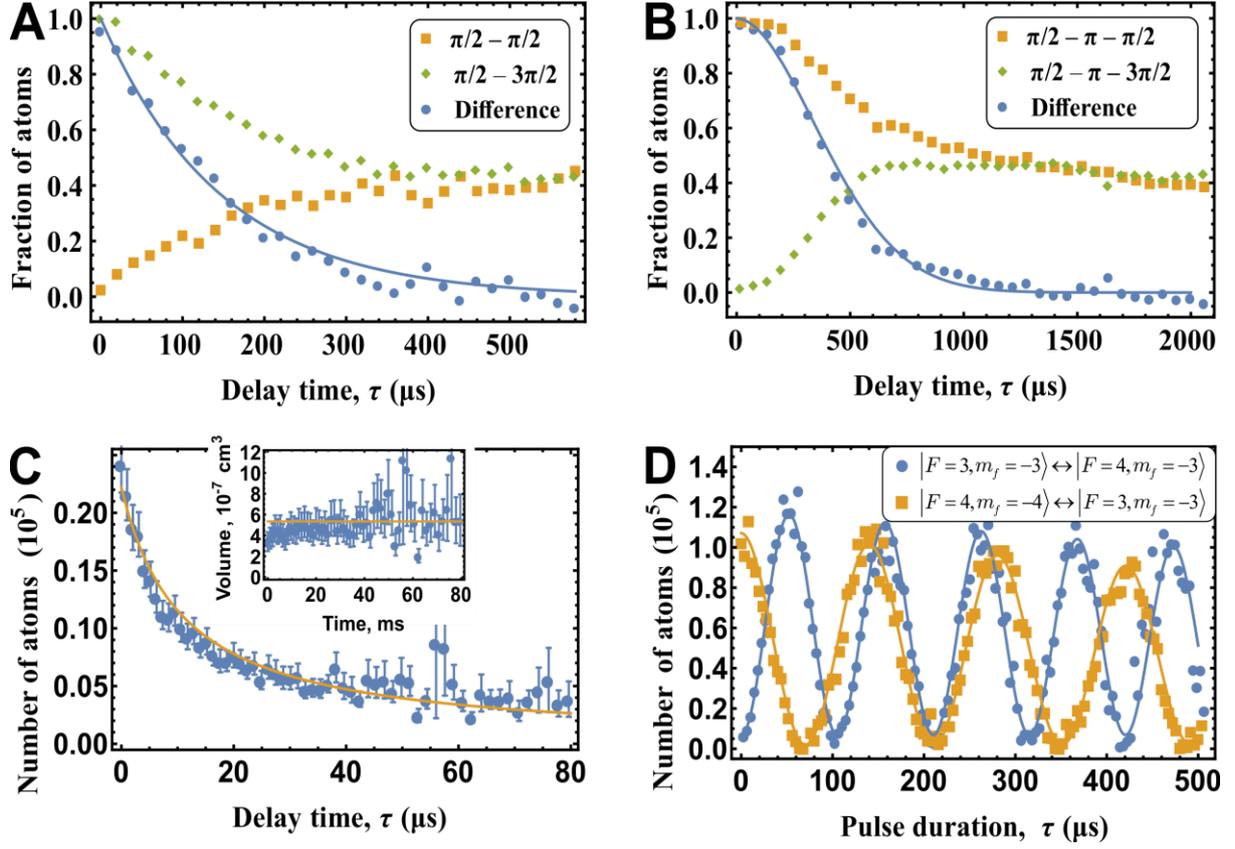

*Figure 4 A – data (dots) and fit of the Ramsey sequence. The fit parameter is $T_2^* = 146$ μs. B – data (dots) and fit of the Hahn echo sequence. The fit parameters are $T_2 = 502$ μs and $\alpha = 1.88$. C – data (dots) and fit (solid line) of $|F=3, m_F=-3\rangle$ state decay under a magnetic field of 5.3 G with only binary loss, $\beta = 2.23 \times 10^{-9}$ cm$^3$/s. The inset indicates the volume of the atomic cloud versus storage time. D – Rabi oscillations in the atomic ensemble for the transition with frequency $f_{43}$ (yellow squares) and $f_{44}$ (blue circles). The solid line represents the fit by (5) with parameters $a = 0.52 \cdot 10^5$, $T_1 = 2513$, $\Omega = 2\pi \cdot 7120$ Hz, and $b = 0.02 \cdot 10^5$ for $f_{43}$ and $a = 0.56 \cdot 10^5$, $T_1 = 2673$, $\Omega = 2\pi \cdot 9513$ Hz, and $b = 0.07 \cdot 10^5$ for $f_{43}$.*

The fidelity of the population transfer from state $|F=4, m_F=-4\rangle$ to state $|F=4, m_F=-3\rangle$ could be estimated from the contrast of the Rabi oscillations. If some of the population after the first transfer pulse remained in the $|F=4, m_F=-4\rangle$ state, then the Rabi oscillation would not reach the 0 level at the minimum of the oscillations. It is clear from Figure 4D that practically all the test population transferred after the first $\pi$-pulse. Nevertheless, if one adds an offset into formula (5), from fitting, one could conclude that $b/(2a+b) = 2\%$. The fit with $b = 0$ is also consistent; therefore, we conclude that no more than 2% remained at level

$|F=4, m_F=-4\rangle$. Similarly, for the Rabi oscillation at frequency $f_{33}$, if any population remained at level $|F=4, m_F=-4\rangle$ or another hyperfine level, then the Rabi oscillation would not have full contrast. If any population remained at level $|F=4, m_F=-4\rangle$, it would again affect the 0 level of the oscillations. The fit gives this level as no more than 6%.

However, since the population at level $|F=4, m_F=-3\rangle$ is originally nearly 0, the population of $|F=3, m_F=-3\rangle$ will only appear in the maximum of the oscillations. Thus, to determine by how much the maximum of the oscillations is different from the original number of atoms, the same transfer sequence was repeated twice: with the MW field turned on and off. The change in the number of atoms detected in the ground state with and without MW was no more than 1%. This number gives by how much the amplitude of the oscillations differs from the one originally present and therefore defines the number of atoms in the $|F=4, m_F=-3\rangle$ state. Thus, at least 93% of the population should be in the $|F=4, m_F=-3\rangle$ state. This does not include any factors related to the purity of the original state; therefore, the actual fidelity of the transfer should be higher. Therefore, the transfer fidelity is over 93%.

## V. CONCLUSION

Population transfer from the $|F=4, m_F=-4\rangle$ state of a thulium-169 cold atomic ensemble to $|F=4, m_F=-3\rangle$ was demonstrated by the consequent application of two $\pi$-pulses between hyperfine levels of the ground state with fidelity $F>93\%$. The transfer was performed using a specially designed MW antenna. The dephasing and coherence times of the intermediate state higher hyperfine components of the thulium atom were measured to be 150 μs and 500 μs, respectively.

The MW antenna design, fabrication and experimental investigation were supported by Russian Science Foundation #19-19-00693. The manipulation of the ultracold thulium atom ground state and all optical experiments were supported by Russian Science Foundation grant #18-12-00266.

## APPENDIX A: ABSORPTION IMAGING

The 410.6 nm transition between the ground state $4f^{13}\left({}^2F^o\right)6s^2\,{}^2F^o_{7/2}\left(I=1/2,F=4\right)$ and $4f^{12}\left({}^3H_5\right)5d_{3/2}6s^2\left(5,3/2\right)\left(I=1/2,F=5\right)$ state was used for absorption imaging. The probe beam had a power of 0.9 mW and a Gaussian profile with a 2.4 mm radius (1/e level). To make the imaging free from complex dynamics between Zeeman manifolds (i.e., with a well-defined absorption cross section), one should use a cyclic transition, for instance, between the lower $m_F=-4$ and upper $m_F=-5$ substates. Thus, the imaging beam should have adequate polarization and be parallel to the DC magnetic field. Here, the directions of the beams were different; therefore, we used the recalibration procedure described in [5]. The calibration factor for the "storage" configuration (Figure 1C) is 1.6 and for "population transfer" is 3.6.

The frequency of the imaging beam was tuned by maximizing the signal from atoms in the $|F=4,m_F=-4\rangle$ state. The detuning of the atoms in the $|F=4,m_F=-3\rangle$ state from the imaging beam is approximately the Zeeman splitting $0.15\Gamma$ ($\Gamma=2\pi\cdot 10$ MHz is the natural width of the imaging transition); thus, atoms in that state can also be efficiently detected. The experimental numbers of atoms measured for $|F=4,m_F=-4\rangle$ and $|F=4,m_F=-3\rangle$ are the same within the experimental error, thus confirming that both states are probed with the same efficiency. In contrast, atoms in $F=3$ states are considerably detuned by $150\Gamma$, making them undetectable for our system, thus allowing us to observe Rabi oscillations between hyperfine sublevels directly from imaging.

## APPENDIX B: G-FACTORS

To calculate g-factors $g_F$, one can use the well-known expression [40]:

$$g_F = g_J \frac{F(F+1)+J(J+1)-I(I+1)}{2F(F+1)}, \quad (6)$$

The g-factor value from [41] $g_J=1.14119$ was used.

## APPENDIX C: CALIBRATION OF THE MW FIELD

The MW radiation for the selected DC magnetic field direction can be represented as the sum of two circularly polarized waves, only one of which addresses the transition. Therefore, formula (3) needs to be rewritten as:

$$B = \frac{\sqrt{2}\hbar\Omega_{Rabi}}{\mu}, \qquad (7)$$

where the factor $\sqrt{2}$ comes from the polarization. Here, $B$ is the magnitude of the AC magnetic field. The magnetic dipole moment $\mu$ for the transition of interest can be calculated as [40]:

$$\mu = \left|\langle \gamma JIFm_F | \mathbb{M} | \gamma JIF'm_F' \rangle\right| = \left|(\gamma JIF | \mathbb{M} | \gamma JIF')\right| \begin{pmatrix} F & 1 & F' \\ -m_F & -1 & m_F - 1 \end{pmatrix}$$

$$\left|(\gamma JIF | \mathbb{M} | \gamma JIF')\right| = g_J \mu_B \sqrt{(I+J+F+1)(J+F-I)(I+F-J)(J+I+1-F)/(4F)} \qquad (8)$$

where $(\gamma F | \mathbb{M} | \gamma F')$ is the matrix element of the transition, $\begin{pmatrix} F & 1 & F' \\ -m_F & -1 & m_F - 1 \end{pmatrix}$ is the 3j-symbol of the transition, $J = 7/2$ is the total angular momentum of the transition, $I = 1/2$ is the nuclear spin, $g_J$ is the Lande g-factor and $\mu_B$ is the Bohr magneton. From (8), one can find the coefficient $\mu = 0.756\mu_B$; therefore, the magnetic field can be found as:

$$B = \frac{\sqrt{2}\hbar\Omega_{Rabi}}{0.756\mu_B} \simeq 1.871 \frac{\hbar\Omega_{Rabi}}{\mu_B} \qquad (9)$$

## APPENDIX D: DECAY FIT

Since the photo of the cloud was taken at $t = 2$ ms of expansion, the initial waist of the cloud can be found as:

$$w_{i0} = \sqrt{w_i^2 - \left(\frac{k_B T}{m_T}t\right)^2} \qquad (10)$$

where $w_i, i \in \{y, z\}$ is the radius at the $1/e$ level of the atomic cloud at 2 ms of expansion, which has a Gaussian density profile, $k_B$ is the Boltzmann constant, $m_T$ is the mass of thulium, and $T$ is the temperature of the cloud. The cloud size along the invisible $x$-axis was calculated through the known aspect ratio of the horizontal ODT beam as $w_{x0}(t) \simeq w_{z0}(t) \cdot \frac{\sigma_x}{\sigma_z}$, where $\sigma_x, \sigma_z$ are the horizontal beam waists at the $1/e$ level. The initial volume of the cloud was almost constant and was calculated as $V = (2\pi)^{3/2} (w_{x0}(t)\, w_{y0}(t)\, w_{z0}(t))$. The systematic uncertainty in the calculation of the volume was estimated as:

$$\frac{\Delta V}{V} = \sqrt{\left(2\frac{\Delta w_z}{w_z}\right)^2 + \left(\frac{\Delta w_y}{w_y}\right)^2 + \left(2\frac{\Delta T}{T}\right)^2}. \tag{11}$$

The uncertainty in temperature measurements was estimated to be 13%. The cloud radius uncertainties $\Delta w_z/w_z = 4\%$ and $\Delta w_y/w_y = 16\%$ lead to a final uncertainty for the initial volume of 32%.

To calculate $\beta$, we fit the data with (4) and then varied the initial volume $V$ from the lowest value of $3.6 \times 10^{-7}$ cm$^3$, which gives $\beta = 1.49 \times 10^{-9}$ cm$^3$/s, to the average value of the endpoint $5.5 \times 10^{-7}$ cm$^3$, with $\beta = 2.28 \times 10^{-9}$ cm$^3$/s. The mean value of the initial value of $5.3 \times 10^{-7}$ cm$^3$, with $\beta = 2.23 \times 10^{-9}$ cm$^3$/s, gives us the bounded rate of $\beta$.

Although the uncertainty in $\beta$ from the fit is small (approximately 6%), the uncertainty in the initial volume made a major contribution to $\beta$ (approximately 32%). Summing up all sources of uncertainty, the final uncertainty for the measured constant of binary collisions is approximately 33%.